\documentstyle[aps,prb,twocolumn,epsf,rotate]{revtex}

\begin{document}

\twocolumn[\hsize\textwidth\columnwidth\hsize\csname
@twocolumnfalse\endcsname

\title{Pauli-Limited Superconductivity in Small Grains}

\author{M. C. B\o nsager and  A.H. MacDonald}
\address{Department of Physics, 
Indiana University, Bloomington, Indiana 47405}
\date{\today}
\maketitle

\begin{abstract}

We report on an exploration of the mean-field phase diagram
for Pauli-limited superconductivity
in small metallic grains.  Emphasis is placed
on the crossover from the ultra-small grain limit where
superconductivity disappears to the bulk thin-film limit as
the single-particle level spacing in the grain decreases.
We find that the maximum Zeeman coupling strength compatible
with superconductivity increases with decreasing grain size,
in spite of a monotonically decreasing condensation energy
per unit volume.

\end{abstract}


\vskip2pc]

\section{Introduction}

Mesoscopic physics may be broadly defined as the study of 
phenomena which depend fundamentally on the finite-size
of a system, even when that size substantially exceeds
characteristic length scales associated with microscopic
degrees of freedom.  By this definition, recent
experiments\cite{tuominen,lafarge,eiles} in which 
strong parity effects are seen in superconducting islands
containing $~\sim 10^9$ electrons highlight the 
robustness of superconductor mesoscopics; 
pairing physics\cite{averin,janko} causes observable differences
between metallic grains 
with $10^9$ electrons and grains with $10^9+1$ electrons.
Recent progress has enabled studies of much smaller systems.
Ralph, Black, and Tinkham\cite{ralph1,ralph2,ralph3} have demonstrated
that it is possible to 
make single-electron transistors with
superconducting islands that are only a few 
nanometers in radius.  These systems contain only $\sim$ 
$10^4$ to $10^5$ electrons and have a mean energy level
spacing $\delta$ which can
be larger than or comparable to $\Delta_0$, the zero-temperature 
superconducting gap in bulk samples. 
As early as in 1959 Anderson observed\cite{anderson} that superconductivity
cannot occur in small grains when the limit $\delta\sim\Delta_0$ is reached. 
Experimental realization of such ultra-small systems has opened
the physics of superconductivity in this regime to experimental
study and inspired substantial theoretical
interest.\cite{agam1,agam2,delft,smith,braun,balian}
Many aspects of the experiments can be qualitatively understood using 
the simplest possible BCS model of an ultra-small grain in which the 
levels are assumed to be equally spaced and pairing is 
assumed to occur only between identical orbitals.\cite{unpub}
In this paper we address the influence of Zeeman
coupling on superconductivity in such a model, emphasizing the 
crossover between the ultra-small grain regime and the 
bulk limit where the Chandrasekhar-Clogston paramagnetic limit, 
$Z < Z_C=\Delta_0/\sqrt{2}$, 
applies.\cite{clogston}  Here $ Z = g \mu_B B/2$ is the Zeeman
coupling strength, $\mu_B$ 
is the Bohr magneton, and $B$ is the magnetic induction. 

For a constant level spacing spectrum, the single particle 
energy levels measured from the Fermi energy are 
$\xi_n=(n-\alpha)\delta$.  Here $n=0,\pm 1,\pm 2,$\ldots,
and parity dependence appears in the quantity $\alpha$ which
has the value $0$ if the 
number of electrons $N$ is odd and $1/2$ if $N$ is even, corresponding  
respectively to chemical potentials pinned at and half-way between
energy levels.  The gap equation 
for a model in which pairing interactions occur only between
identical orbitals differs from its BCS theory counterpart 
only\cite{delft,unpub} in the discreteness of the quasiparticle 
level spectrum:
\begin{equation}\label{gapeqn}
\frac{1}{\lambda}=\delta\sum_{n=1}^{M} 
\frac{1-f(E_n+Z)-f(E_n-Z)}{E_n},
\end{equation}
where $E_n=\sqrt{\Delta^2+\xi_n^2}$, 
$\lambda$ is the dimensionless coupling constant and $\Delta$ is 
determined by solving these equations. The upper limit on
this discrete sum comes from the energy cutoff used in BCS theory 
to model retarded attractive interactions
and can be expressed in terms of $\Delta_0$ using the bulk solution of the
zero temperature gap equation 
\begin{equation}
\Delta_0 = 2 M \delta \exp(-1/\lambda).
\label{eq:upperlimit}
\end{equation}  
Since, at available fields, the magnetic flux through an ultra-small grains
will typically be much smaller than $\Phi_0 = \hbar c / 2e$,
coupling to orbital
degrees of freedom\cite{tinkhamdegennes} can normally be ignored.
The electron spins still couple to the magnetic field $B$,
however, splitting the single-particle 
energies, $\xi_{n} \to (n-\alpha) \delta \pm Z$.

For small superconducting particles it is essential that the 
occupation probabilities, $f$, in Eq. (\ref{gapeqn}) be
calculated in ensembles including states with only
even or odd numbers of particles.\cite{janko}
These differ from Fermi occupation probabilities only for 
levels close to the chemical potential and 
only for temperatures $k_BT\lesssim\delta$; at T=0 the even restriction
has no effect and the odd restriction has only the effect of removing the 
orbital at the Fermi energy, which cannot be paired, from the gap equation.
This model was first studied by von Delft et al.\cite{delft} 
to calculate the dependence of $\Delta(T)$ on $\delta$. They found that 
$\Delta(T)$ remains close to its bulk value until $\delta$
is close to a critical value $\delta_c$ which is parity dependent:
$\delta_c^{\rm odd}/\Delta_0=\frac{1}{2}e^{\gamma}\simeq 0.89$ and 
$\delta_c^{\rm even}/\Delta_0=2e^{\gamma}\simeq 3.56$. Here 
$\gamma=0.577215...$ is Euler's constant. 
As $\delta\rightarrow\delta_c$ from below, the critical temperature and the 
zero temperature gap both tend to zero. 
Later Braun et al.\cite{braun} and Balian et al.\cite{balian} extended 
this work to the case of finite Zeeman coupling. Ref.\onlinecite{braun} 
concentrated on comparison between theory and the 
experiments by Ralph et al.\cite{ralph3}, finding good
qualitative agreement.  Ref.\onlinecite{balian} concentrated 
on the influence of  
using parity dependent distribution functions at finite
temperature, predicting significant qualitative effects
for such quantities as the superconducting gap and the 
magnetization.  Here we explore the full $T-Z$ phase diagram, 
examining dependences on parity and $\delta$ and focussing on 
the evolution of the $T-Z$ phase diagram toward its 
bulk limit result as $\delta$ decreases.

\section{Phase Diagrams} 

The fundamental equation on which our calculations are based 
is the following coupling-constant-integration expression
for the free-energy difference between superconducting and 
normal states:\cite{fw}
\begin{eqnarray}\label{freediff}
\Omega_s-\Omega_n &=& 
\frac{\Delta^2}{\lambda\delta}\nonumber\\
&&\!\!\!\!\!\!\!\!\!\!\!\!\!\!\!\!\!\!\!
-2\sum_{n=1}^M \int_0^{\Delta}
d\Delta'\Delta'
\frac{1-f(E'_n+Z)-f(E'_n-Z)}{E'_n}.
\end{eqnarray}
In Eq. (\ref{freediff}) $E'_n=\sqrt{\xi_n^2+\Delta'^2}$.
Since, in the second term on the right hand side 
of Eq. (\ref{freediff}), $\Delta$ appears only in the upper limit 
of the coupling-constant integration, it is easy to verify that the 
gap equation, Eq. (\ref{gapeqn}), is satisfied at extrema of the 
condensation energy $\Omega_s-\Omega_n$.
A sufficient condition for
superconductivity in this model is that
$\partial_{\Delta^2}(\Omega_s-\Omega_n)|_{\Delta^2=0}$ be negative;
this derivative changes sign along the surface in $(Z,T,\delta)$ 
space where the linearized ($\Delta^2 \to 0$) gap equation is 
satisfied.  When the phase transition is continuous,  
this surface, $Z_2(T,\delta)$, is the boundary of the superconducting 
region. If, on the other hand, 
$\partial_{\Delta^2}(\Omega_s-\Omega_n)|_{\Delta^2=0}$ is 
positive the grain may still be in a superconducting phase if 
$\Omega_s-\Omega_n$ is negative at a finite value of $\Delta$. In 
that case the phase boundary is of first order and the corresponding 
critical value of $Z$ is denoted $Z_1(T,\delta)$. 

Below we determine the functions $Z_2(T,\delta)$ and $Z_1(T,\delta)$, 
and thereby show how the phase diagram for bulk systems is generalized 
to systems with finite level spacing.

\subsection{Normal state instability at $T=0$}

It is instructive to start by considering $Z_2(T=0,\delta)$ which can
be evaluated analytically by solving the linearized gap equation 
(\ref{gapeqn}). To that end 
we note that at $T=0$ the factor $[1-f(\xi_n+Z)-f(\xi_n-Z)]$ 
in Eq. (\ref{gapeqn}) equals zero if 
$\xi_n-Z$ is negative and otherwise equals one.
Hence, as $Z$ is increased an additional pair 
of states ($n\uparrow,n\downarrow$), is blocked from pairing every 
time $\alpha+Z/\delta$ passes through an integer value. 
It follows that 
\begin{equation}\label{z2eqn}
Z_2(T=0,\delta) = (m-\alpha)\delta
\end{equation}
where $m$ is the largest integer for which 
\begin{equation}\label{ineq}
\sum_{n=m}^{M}\frac{1}{n-\alpha} = \psi(M+1-\alpha)
 - \psi(m-\alpha) >  \frac{1}{\lambda}.
\end{equation}
Here $\psi(x)$ is Euler's $\psi$ function.  
Using $\psi(x) \sim {\rm ln} (x)$ for large arguments 
and replacing $M$ using Eq. (\ref{eq:upperlimit}) we find that 
$m$ is the largest integer for which 
\begin{equation}\label{ineq2}
\delta/\Delta_0 < \frac{1}{2} \exp[ - \psi(m-\alpha)].  
\end{equation}
The resulting $Z_2(T=0,\delta)$ is plotted in Fig. \ref{zzeven} and 
Fig. \ref{zzodd} for even and odd $N$, respectively. 
In both cases the bulk value\cite{sarma} $Z_2(T=0,\delta\rightarrow 0)=1/2$ 
is recovered as can be verified by letting 
$m$ become large in Eqs. (\ref{z2eqn}) and (\ref{ineq2}). 
Rather than decresing steadily, $Z_2(T=0,\delta)$ 
oscillates around its $\delta\rightarrow 0$ limit with increasing $\delta$. 
The maximum values of $Z_2(T=0,\delta)$ 
actually occur for $\delta\rightarrow\delta_c$ ($m=1$); 
$Z_2^{\rm max}/\Delta_0=e^{\gamma}\simeq 1.78$ for even $N$ and 
$Z_2^{\rm max}/\Delta_0=e^{\gamma}/2 \simeq 0.89$ for odd $N$.
Both of these values exceed $Z_{C}/\Delta_0=1/\sqrt{2}$, 
and this analytic result thus establishes that the
Chandrasekhar-Clogston limit\cite{clogston} can be exceeded 
in the ultra-small particle
limit and, as we see below, also over a broad range of 
small particle sizes.  The decrease of $\Delta$ and of 
mean-field-theory critical temperatures with particle size 
is not accompanied by a corresponding decrease in the 
maximum allowed Zeeman coupling strength. 

\subsection{First-order Transition Phase Boundary at $T=0$} 

When $\Delta$ is finite, the pair-breaking condition 
($Z>\sqrt{\xi_n^2+\Delta^2}$) is not satisfied until larger values
of $Z$ are reached compared to the $\Delta=0$ case.
Hence, at sufficiently low temperatures, 
states with finite $\Delta$ are favored over states with 
$\Delta\rightarrow 0$, causing the superconductor-normal transition to 
be of first order.  This physics is much the same 
at finite $\delta$ and in the $\delta \to 0$ bulk thin film
limit.

\begin{figure}
\epsfxsize=6.6cm
\rotate[r]{\epsfbox{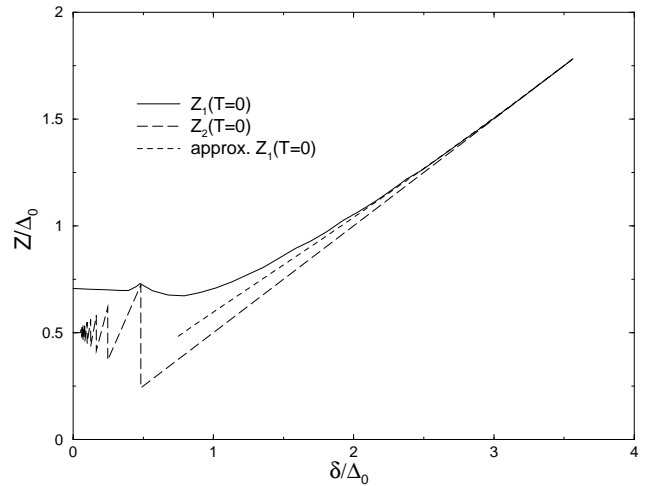}}
\vspace{0.5cm}
\caption[]{
The zero temperature limit of $Z_1$ (solid line) and $Z_2$ (long-dashed line) 
as a function of $\delta$ for an even number of electrons. The 
dashed line is the approximate expression (\ref{z1approx}) for $Z_1$
derived in the text. $Z_2$ can be evaluated analytically as
explained in the text. 
The discontinuities in this curve occur at 
$\delta/\Delta_0 = 0.4821,\ 0.2475,\ 0.1659,\ 0.1247,\ ...$ and 
the corresponding minima and maxima are given by Eq. (\ref{z2eqn}). 
The function $Z_1(T=0,\delta)$ is continous but it has discontinuities 
in its first derivative at a series of $\delta$. This fact is 
further illustrated in Fig. \ref{zzodd}.
}
\label{zzeven}
\end{figure}

\begin{figure}
\epsfxsize=6.6cm
\rotate[r]{\epsfbox{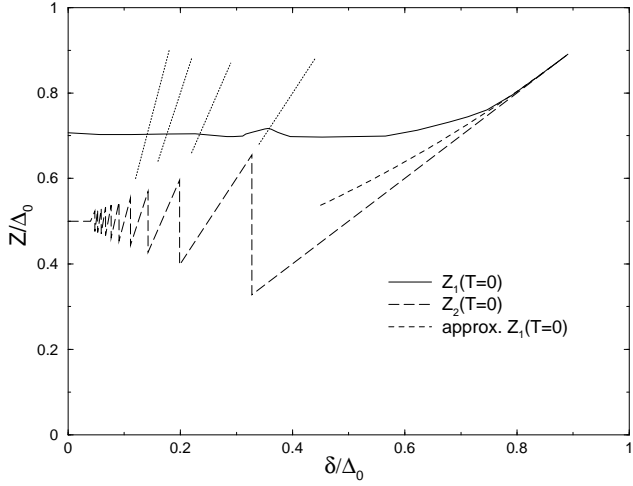}}
\vspace{0.5cm}
\caption[]{
As in Fig.~\ref{zzeven} but for an odd number of electrons. 
The discontinuities in $Z_2$ occur at 
$\delta/\Delta_0 = 0.3276,\ 0.1987,\ 0.1424,\ 0.1109,\ ...$.
As it is the case for even $N$ (see Fig. \ref{zzeven}), the 
function $Z_1(T=0,\delta)$ has discontinuities in its first 
derivative at a series of $\delta$ given by 
$Z_1(\delta)=(n-\alpha)\delta$, $n=2,3,4,\ ...$, i.e. where 
extrapolations of $Z_2$ (indicated by dotted lines) cross 
$Z_1$. These cusps are a consequence of discontinuities of 
the first derivative of the normal state energy with respect 
to $Z$. In most cases, the cusps are too small see in 
this figure, however.
}
\label{zzodd}
\end{figure}

We consider first the ultra-small grain limit. 
At $T=0$ the integral in Eq. (\ref{freediff}) can be evaluated analytically. 
\begin{eqnarray}
\Omega_s-\Omega_n &=& \frac{\Delta^2}{\lambda\delta}
-2\sum_{n=1}^{M}\{ {\rm max}(E_n,Z)-{\rm max}(\xi_n,Z) \} \nonumber\\
&\simeq &
\frac{\Delta^2}{\delta}\ln(\delta/\delta_c)
+\frac{\Delta^4}{4\delta^3}\zeta(3,\alpha) + 2(Z-Z_2).
\label{cond}
\end{eqnarray}
The first form for the right hand side of Eq. (\ref{cond}) is exact 
whereas the second form only 
applies in the ultra-small grain regime: 
$\xi_2 > Z > \xi_1$, $E_1 > Z$ and $\Delta/\delta \ll 1$.
In Eq. (\ref{cond}), $\zeta$ is Riemann's Zeta Function; 
$\zeta(3,0)\simeq 1.2021$ and $\zeta(3,1/2)\simeq 8.4144$.
Minimizing $\Omega_s-\Omega_n$ we find that 
$\Delta^2=2\delta^2 \ln(\delta_c/\delta)/\zeta(3,\alpha)$ 
and 
\begin{equation}\label{z1approx}
Z_1(T=0,\delta)=
Z_2(T=0,\delta)+\delta\frac{[\ln(\delta_c/\delta)]^2}{2\zeta(3,\alpha)}
\end{equation}
for $\delta\rightarrow\delta_c$. This expression is plotted together with 
exact numerical evaluations of $Z_1(T=0,\delta)$ in Figs. \ref{zzeven} and 
\ref{zzodd} for even and odd $N$ respectively. 
$Z_1(T=0,\delta)$ and $Z_2(T=0,\delta)$ become equal
only as $\delta \to \delta_c$.

\begin{figure}
\epsfxsize=6.6cm
\rotate[r]{\epsfbox{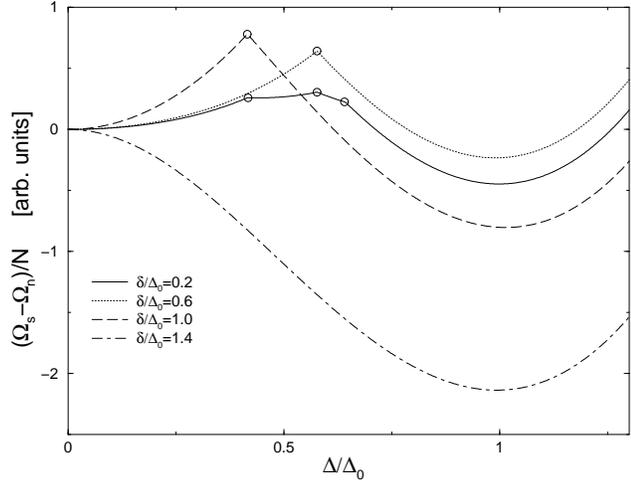}}
\vspace{0.5cm}
\caption[]{
The dependence of the free energy difference per particle 
$(\Omega_s-\Omega_n)/N$ on $\Delta$ at $(T=0,Z=0.65\Delta_0)$
for four different values of level spacing $\delta$ for an 
even number of electrons. The circles indicates cusps in the 
curves. 
The case $\delta/\Delta_0=1.4$ is special in the sense that 
it corresponds to Zeeman field below $Z_2(\delta)$ 
(see Fig. \ref{zzeven}). For $Z>Z_2(1.4\Delta_0)$ this curve 
would also have a positive slope at $\Delta=0$.
}
\label{condeng}
\end{figure}

We notice in Fig. \ref{zzeven} that the exact numerical evaluation  
recovers the familiar Chandrasekhar-Clogston result in the 
bulk limit: $Z_1(T=0,\delta\rightarrow 0)=1/\sqrt{2}\Delta_0$ 
is recovered for both even $N$ and odd $N$. 
The exact numerical results 
for $Z_1(T=0,\delta)$ were obtained by evaluating $\Omega_s-\Omega_n$ 
as a function of $\Delta$ and locating zero-crossings of its minimum 
as $Z$ is increased. 
In Fig. \ref{condeng} we show the dependence of 
$(\Omega_s-\Omega_n)/N$ on $\Delta$ for four different 
values of $\delta$ at $(T=0,Z=0.65\Delta_0)$ for an even number 
of electrons. 
For $\delta/\Delta_0=1.4$ this corresponds to $Z<Z_2(\delta)$, but for 
the three lowest values of $\delta$, $Z_2(\delta)<Z<Z_1(\delta)$ 
and the slope at $\Delta=0$ is therefore positive.
$\Omega_s-\Omega_n$ nevertheless becomes negative at a finite
value of $\Delta$,  
and in all cases the optimal value of $\Delta$ (for which 
$\Omega_s-\Omega_n$ has its minimum) is very close to the 
bulk value $\Delta_0$.  As shown by 
von Delft {\it et al.},\cite{delft}
only when $\delta$ is very close to $\delta_c$ 
does the optimal value diminish significantly.
It is furthermore worth noticing that, as can easily be concluded 
from Eq. (\ref{cond}), the optimal value of $\Delta$ 
is independent of $Z$, for $Z < Z_1$. 

Each time the number of Pauli blocked pairs decreases with 
increasing $\Delta$, the free energy curve has an
upward-pointing cusp as exemplified in Fig. \ref{condeng} by the 
three smallest values of $\delta$. The cusps are spaced more closely at 
smaller values of $\delta$, but the overall envelopes of the three 
curves are rather similar.  
(The fourth curve has no cusps because neither normal nor
superconducting states are spin-polairzed.) 
In a model with constant energy level 
spacings, as employed here, the minimum of $\Omega_s-\Omega_n$ 
always occurs with a minimum of blocked pairs, i.e. in the 
superconducting state the grain will have spin zero for even $N$ 
and spin 1/2 for odd $N$. 

\subsection{Full $T-Z$ phase diagrams}

\begin{figure}
\epsfxsize=6.6cm
\rotate[r]{\epsfbox{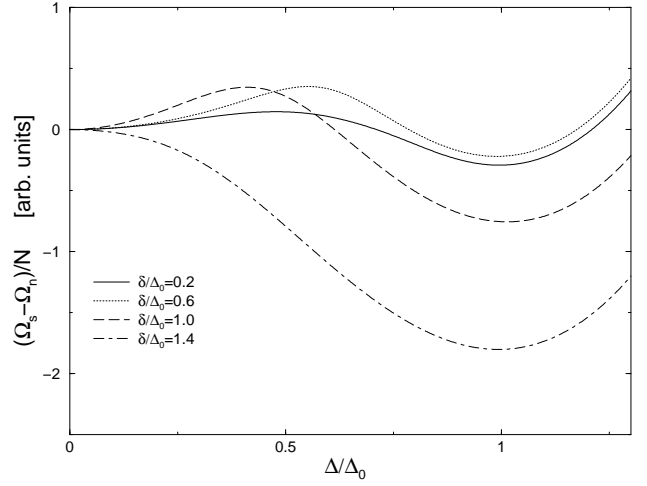}}
\vspace{0.5cm}
\caption[]{
The same as in Fig. \ref{condeng} but at finite temperature 
$T/T_{c0}=0.2$, where $T_{c0}$ is the bulk critical temperature 
at $Z=0$.
}
\label{condengfinitet}
\end{figure}

\begin{figure}
\epsfxsize=6.6cm
\rotate[r]{\epsfbox{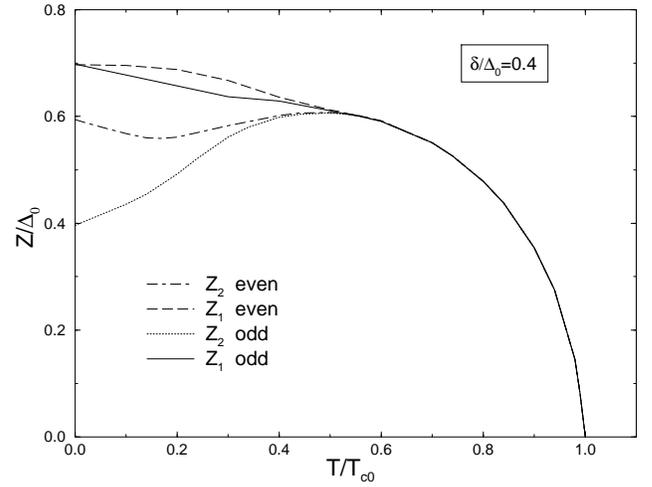}}
\vspace{0.5cm}
\caption[]{
The functions $Z_1(T)$ and $Z_2(T)$ for both even and odd numbers 
of electrons. This figure is for $\delta/\Delta_0=0.4$. 
Temperatures are expressed in terms of the bulk critical temperature $T_{c0}$.
For $\delta/\Delta_0=0.4$ $Z_1(T=0)/\Delta_0$ is already close to its 
bulk value, $1/\sqrt{2}$. 
For both even and odd number of electrons the transition becomes first order
near $T/T_{c0}=0.56$, close to the corresponding bulk value. 
$Z_2$ shows the largest deviation from the bulk limit. For 
odd number of particles $Z_2(T=0,\delta)<Z_2(T=0,\delta\rightarrow 0)$ 
while for even number of particles
$Z_2(T=0,\delta)>Z_2(T=0,\delta\rightarrow 0)$.
}
\label{trefem}
\end{figure}

\begin{figure}
\epsfxsize=6.6cm
\rotate[r]{\epsfbox{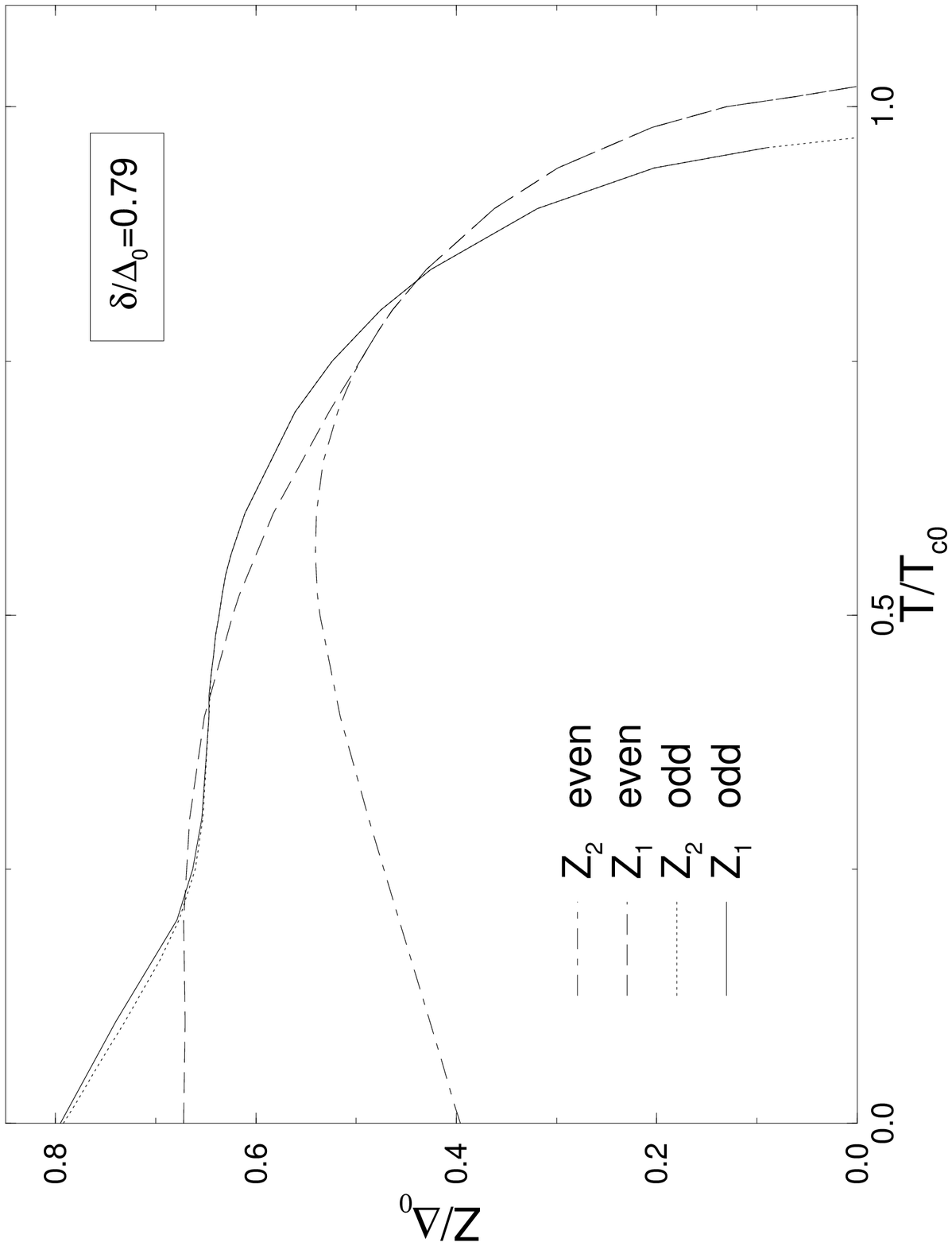}}
\vspace{0.5cm}
\caption[]{
The functions $Z_1(T)$ and $Z_2(T)$ for both even and odd numbers 
of electrons. This figure is for $\delta/\Delta_0=0.79$, 
a value sufficiently large to yield phase diagrams which  
differ substantially from their bulk counterparts. 
The transition becomes first order at $T/T_{c0}=0.76$ for even
$N$ and at $T/T_{c0}=0.44$ for odd $N$. 
This illustration also reflects the dependence of the 
critical temperature $\delta$ and parity discussed by by
von Delft {\it et al.}\protect\cite{delft}. For odd $N$ $T_c$ decreases 
monotonically with $\delta$ whereas for even $N$ it increases up to about 
$\delta/\Delta_0\sim 2$ before it decreases. 
}
\label{syv}
\end{figure}

At finite temperature both of the functions $Z_1(T,\delta)$ and 
$Z_2(T,\delta)$ are found by numerically analyzing 
$\Omega_s-\Omega_n$ as a function of $\Delta$. In 
Fig. \ref{condengfinitet} we show the finite temperature version 
of the plots in Fig. \ref{condeng}. The cusps are thermally 
broadened and the condensation energies have diminished for 
all values of $Z$. 
The full $T$-$Z$ phase diagram depends on both $\delta$ 
and electron-number parity. Examples at representative 
intermediate values of $\delta$ are presented in 
Figs. \ref{trefem} and \ref{syv}. 
The general picture is similar to the bulk case,\cite{sarma}
even as $\delta/\Delta $ approaches one.
At low temperature $Z_1(T,\delta)>Z_2(T,\delta)$ and 
the transition to the normal state is first order. At the transition 
$\Delta$ then drops abruptly from its $Z=0$ value,
$\Delta(T,\delta,Z=0)$, to zero. 
At higher temperatures, $Z_1(T,\delta)=Z_2(T,\delta)$ and the transition 
is continuous. In that case, the free energy curves 
do not cross zero at finite $\Delta$ if their initial slope is 
positive.  

Whereas $Z_1(T\rightarrow 0,\delta)$ only differs from its bulk value 
in the ultra-small limit, $Z_2(T\rightarrow 0,\delta)$ has a 
significant $\delta$-dependence even for intermediate grain sizes. 
Although $Z_2$ doesn't correspond to a phase transition at low 
temperatures, it does have physical meaning as a supercooling curve, 
and has been successfully 
addressed experimentally in thin films\cite{meservey} by measuring 
enhanced fluctuations in the neighborhood of $Z_2$.

\section{Discussion} 

Level spacings near the chemical potential in a real ultra-small grain will
fluctuate\cite{smith} around the mean-value used in our idealized 
model.  For an even number of electrons $\mu(T=0,Z=0)$
will fall between\cite{fn} two energy levels $\epsilon_a$ and $\epsilon_{a-1}$. 
For an odd number $\mu(T=0,Z=0)$ will fall close to\cite{fn} a level 
$\epsilon_a$. Since our results for $Z_2(T=0,\delta\rightarrow\delta_c)$
depend primarily on Pauli blocking of the first pair of
levels, the main consequence of using a realistic spectrum 
are captured by identifying $\delta$ with 
$\epsilon_a-\epsilon_{a-1}$ for even $N$ and 
$(\epsilon_{a+1}-\epsilon_{a-1})/2$ for odd $N$.
As a consequence, like 
minimum grain sizes,\cite{smith} maximum Zeeman energies 
will have a broad distribution.  Consequently,
it is difficult to compare directly with specific data.
Nevertheless our work appears to shed some light on the interpretation
of both recent and older experiments. 

In a pioneering early experiment, Giaever and Zeller\cite{giaever}
found a violation of the Chandrasekhar-Clogston limit which, to our knowledge, 
is still not fully explained.  These authors 
studied tunneling through an ensemble of Sn grains with 
a narrow size distribution.  Interpreting their measurements using a Coulomb 
blockade picture, they concluded that most 
grains retained a superconducting gap up to magnetic fields that exceeded 
the Chandrasekhar-Clogston limit. 
The more recent experiments by Ralph, Black, and Tinkham\cite{ralph3}
also appear to find 
that the Chandrasekhar-Clogston limit can be exceeded. 
In the later tunneling experiments, a 
quasiparticle gap, is observed to decrease linearly with the Zeeman
coupling strength $Z$.
The linear dependence arises from the Zeeman splitting of quasiparticle
energies and is consistent with the mean-field theory employed in 
this paper.\cite{braun} The linear decrease 
continues to Zeeman fields that exceed the Chandrasekhar-Clogston limit
without the discontinuous drop which would be expected if 
$\Delta$ dropped abruptly to zero.  However, this observation 
is made at a value of $\Delta_0/\delta$ which is smaller than
those for which the Chandrasekhar-Clogston limit is exceeded in a model with
equally spaced energy levels.  The experiment could be explained 
by assumming that this particular sample happens to have a relatively
large energy spacing at the Fermi energy.

The results presented in this paper are based on mean-field theory, and 
a few cautionary remarks concerning its validity are in order. For 
$T=0$ and $Z\leq Z_2$, the mean-field condensation energy for  
$\delta\rightarrow\delta_c$ becomes microscopic:
$\Omega_s-\Omega_n \to -\delta [\ln(\delta_c/\delta]^2/\zeta(3,\alpha)$.
Thermal fluctuations will therefore be important unless the temperature
is well below the bulk critical temperature and quantum fluctuations\cite{fluc} 
will be increasingly important as $\delta_c$ is approached.
Clearly the ultrasmall grain
in this regime will not exhibit anything approaching a true phase 
transition between normal and superconducting states.  The phase boundaries
found in mean-field-theory should be regarded as estimates for 
the localtions of crossovers which become more gradual as $\delta$ 
increases.
 
This work was supported in part by NSF under Grant numbers DMR9714055 
and PHY9407194, and in part by the Danish Research Academy. 
The authors would like to thank ITP, Santa Barbara for its 
hospitality. Discussions with F. Braun, J. von 
Delft, D. Ralph, M. Tinkham, and A. D. Zaikin are gratefully acknowledged.  



\begin{references}
\bibitem{tuominen} M. T. Tuominen, J. M. Hergenrother, T. S. Tighe, and 
M. Tinkham, Phys. Rev. Lett. {\bf 69}, 1997 (1992). 
\bibitem{lafarge} P. Lafarge, P. Joyez, D. Esteve, C. Urbina, and M. H. Devoret,
Phys. Rev. Lett. {\bf 70}, 994 (1993).
\bibitem{eiles} T. M. Eiles, J. M. Martinis, and M. H. Devoret, 
Phys. Rev. Lett. {\bf 70}, 1862 (1993).
\bibitem{averin} D. V.Averin and Yu. V. Nazarov, Phys. Rev. Lett. {\bf 69}.
1993 (1992)
\bibitem{janko} B. Jank\'{o}, A. Smith, and V. Ambegaokar, Phys. Rev. B, 
{\bf 50}, 1152 (1994).
\bibitem{ralph1} D. C. Ralph, C. T. Black, and M. Tinkham, Phys. Rev.. Lett. 
{\bf 74}, 3241 (1995).
\bibitem{ralph2} C. T. Black, D. C. Ralph, and M. Tinkham, 
Phys. Rev. Lett. {\bf 76}, 688 (1996).
\bibitem{ralph3} D. C. Ralph, C. T. Black, and M. Tinkham, 
Phys. Rev. Lett. {\bf 78}, 4087 (1997).
\bibitem{anderson} P. W. Anderson, J. Phys. Chem. Solids {\bf 11}, 26 (1959).
\bibitem{agam1} O. Agam, N. S. Wingreen, B. L. Altshuler, D. C. Ralph, 
and M. Tinkham, Phys. Rev. Lett. {\bf 78}, 1956 (1997).
\bibitem{agam2} O. Agam and I. Aleiner, Phys. Rev. B {\bf 56}, R5759 (1997).
\bibitem{delft} J. von Delft, A. D. Zaikin, D. S. Golubev, and W. Tichy, 
Phys. Rev. Lett. {\bf 77}, 3189 (1996).
\bibitem{smith} R. A. Smith and V. Ambegaokar, Phys. Rev. Lett. {\bf 77}, 
4962 (1996).
\bibitem{braun} F. Braun, J. von Delft, D. C. Ralph, and M. Tinkham, 
Phys. Rev. Lett. {\bf 79}, 921 (1997). F. Braun and J. von Delft, 
cond-mat/9801170..
\bibitem{balian} R. Balian, H. Flocard, and M. V\'{e}n\'{e}roni, 
nuc-th/9706041 and cond-mat/9802006.
\bibitem{unpub} A critical examination of the assumption 
that pairing occurs only between identical orbitals will be presented elsewhere.
M. C. B{\o}nsager and A. H. MacDonald, in preparation, to be submitted
to Physical Review B. 
\bibitem{clogston} A. M. Clogston, Phys. Rev. Lett. {\bf 9}, 266 (1962);
B. S. Chandrasekhar, Appl. Phys. Lett. {\bf 1}, 7 (1962).
\bibitem{tinkhamdegennes} P.-G. de Gennes and M. Tinkham,
Physica {\bf 1}, 107 (1964).
\bibitem{fw} A. L. Fetter and J. D. Walecka, {\it Quantum Theory of 
Many-Particle Systems}, McGraw-Hill, New York, 1971.
\bibitem{sarma} G. Sarma, J. Phys. Chem. Solids {\bf 24}, 1029 (1963);
K. Maki, and T. Tsuneto, Prog. Theor. Phys. {\bf 31}, 945 (1964).
\bibitem{meservey} R. Meservey and P. M. Tedrow, Phys. Rep. {\bf 238}, 
173 (1994).
\bibitem{fn} For a constant energy level spectrum $\mu$ is independent of $T$ and $Z$
and falls exactly on a level (odd $N$) or exactly between two levels (even $N$). 
This is not the case for a random spectrum. Rather, $\mu$ will depend on both 
$T$ and $Z$ and it will not fall exactly on or between levels. 
M.C. B{\o}nsager and A.H. MacDonald, unpublished.
\bibitem{giaever} I. Giaever and H. R. Zeller, Phys. Rev. Lett. {\bf 20}, 
1504 (1968); H. R. Zeller and I. Giaever, Phys. Rev. {\bf 181}, 789 (1969).
\bibitem{fluc} K. A. Matveev and A. I. Larkin, Phys. Rev. Lett. {\bf 78}, 
3749 (1997); A. Mastellone, G. Falci, and R. Fazio, Phys. Rev. Lett. {\bf 80}, 
4542 (1998); S. D. Berger and B. I. Halperin, Phys. Rev. B {\bf 58}, 5213 (1998).

\end{references}
\end{document}